\documentclass{iaus278}
\usepackage{graphicx}
\pubyear{2011}
\volume{278}  
\pagerange{}
\jname{Oxford IX: International Symposium on\\ Archaeoastronomy \& Astronomy in Culture}
\editors{Clive Ruggles, edt.}

\title[Australian Cultural Astronomy]                   
{``Bridging the Gap" through Australian Cultural Astronomy}

\author[Duane W. Hamacher \& Ray P. Norris] 
{Duane W. Hamacher \and Ray P. Norris}        

\affiliation{Department of Indigenous Studies, Macquarie University, NSW, 2109, Australia\\ 
                 email: {\tt duane.hamacher@mq.edu.au}\\[\affilskip]}

\begin{document}
\maketitle

\begin{abstract}
For more than 50,000 years, Indigenous Australians have incorporated celestial events into their oral traditions and used the motions of celestial bodies for navigation, time--keeping, food economics, and social structure.  In this paper, we explore the ways in which Aboriginal people made careful observations of the sky, measurements of celestial bodies, and incorporated astronomical events into complex oral traditions by searching for written records of time--keeping using celestial bodies, the use of rising and setting stars as indicators of special events, recorded observations of variable stars, the solar cycle, and lunar phases (including ocean tides and eclipses) in oral tradition, as well as astronomical measurements of the equinox, solstice, and cardinal points.

\end{abstract}

\firstsection

\section{Introduction}
An accumulation of evidence has shown that Aboriginal Australians were careful observers of the night sky and that celestial knowledge played a major role in the culture, social structure, and oral traditions of the hundreds of distinct Aboriginal groups, each with a distinct language and culture, that existed prior to British colonisation (e.g. Stanbridge, 1858; Griffin, 1923; Maegraith, 1932; Mountford, 1958; Clarke, 1997; Johnson, 1998; Haynes, 2000; Cairns \& Harney, 2003; Fredrick, 2008; Norris \& Norris, 2009; Norris \& Hamacher, 2009), something Aboriginal people themselves have long known but is only beginning to receive long--overdue acknowledgement by mainstream Australians.  This knowledge included an understanding that celestial phenomena correlated to terrestrial events, such as the passage of time, the changing of seasons, the emergence of particular food sources, the timing of ocean tides, and the nature of transient celestial phenomena, such as comets, eclipses, meteors, and cosmic impacts (e.g. Hamacher \& Norris 2009, 2010, 2011a, 2011b, Hamacher \& Frew, 2010).  Aboriginal people used the sky for marriage and totem classes and as cultural mnemonics (Johnson, 1998).  This knowledge was passed through successive generations via oral tradition, dance, ceremony, and various artistic forms.  Much of this knowledge was restricted to particular genders, totems, or was dependant on the initiation of that individual into the higher ranks of the community.

It has been claimed that Aboriginal people ``made no measurements of space and time, nor did they engage in even the most elementary of mathematical calculations" (Haynes, 2000:54).  As recently as the 1980s, Blake (1981) stated that ``no Australian Aboriginal language has a word for a number higher than four,"  despite well--documented Aboriginal number systems (e.g. McRoberts, 1990; Tully, 1997).  If indeed Aboriginal people were incapable of counting past four, then it would seem unlikely that they ``measured things" or observed celestial phenomena.  Such misconceptions are an obstacle to research in this area. 

While some Aboriginal groups were badly damaged by British colonisation, other communities, especially those in Arnhem Land, still live fairly ``traditional" lifestyles, where the traditional language is spoken, and ceremonies, laws, and artistic forms are strong and vibrant.  However, in other regions, evidence of once strong Aboriginal cultures is confined to historical accounts or archaeological sites, largely due to the damaging effects of colonisation.

In this paper, we highlight examples of how Aboriginal people made use of celestial phenomena for calendric, navigational, and cultural purposes and show that Aboriginal people were careful observers of the night sky.  We show that they noted the changing brightness of particular stars, and the complex motions of the sun and moon, and  oriented stone arrangements to cardinal directions, and to locations of astronomical phenomena.

\section{Timekeeping \& Written Records}

\begin{figure}[h]
\begin{center}
 \includegraphics[width=6cm]{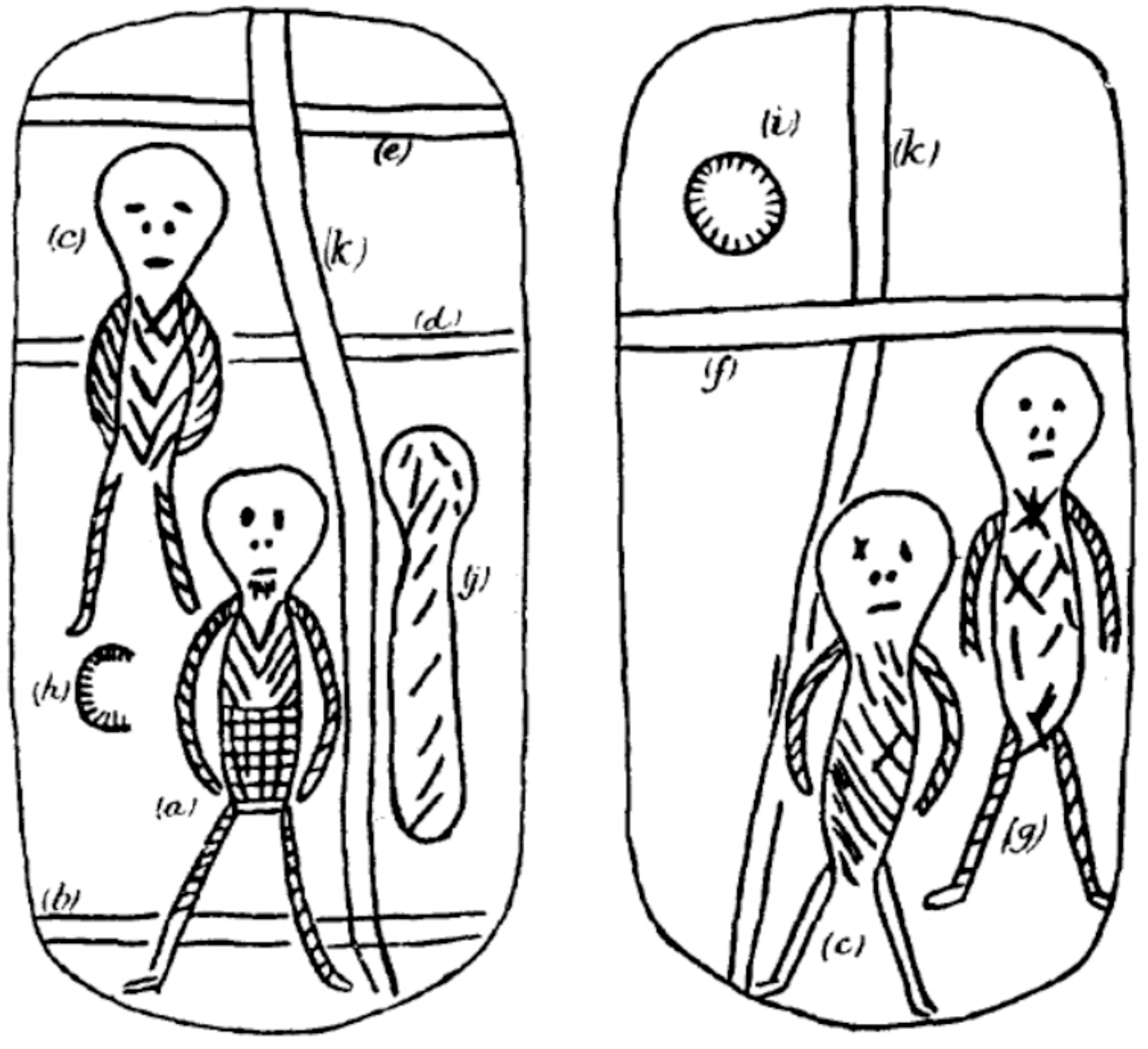} 
 \caption{A message stick, taken from Mathews (1897:292), depicting information including time, denoted by the phase of the moon.}
   \label{fig:message_stick}
\end{center}
\end{figure}

There are a number of ways celestial objects can be used to record the passage of time.  In Aboriginal cultures, the moon is widely used.  Hahn (1964:130) notes that Aboriginal people of the Hahndorf area in the Adelaide Hills were observed making notches in their digging sticks upon the appearance of each New Moon to mark their own age.  Message sticks consist of pictograms used to communicate particular information to distant communities, which Howitt (1898:314) called ``Blackfellow's letters".  For example, Mathews (1897:293) explains how the pictograms on a message stick (Fig. \ref{fig:message_stick}) represent information about the location and time of a corroboree to be held in the future.  The message stick states that ``Nanee (a) sent the message from the Bokhara river (b), by the hand of Imball (c), via the Birie (d), the Culgoa (e), and Cudnappa (f) rivers, to Belay (g); that the stick was dispatched at new moon (h), and Belay and his tribe are expected to be at Cudnappa river (f) at full moon (i); (j) represents a corroboree ground, and Belay understands from it that Nanee and his tribe are corroboreeing at the Bokhara river, which is their taorai, and, further, that on the meeting of the two tribes at full moon on the Cudnappa river a big corroboree will be held."

This particular message stick reveals pictograms representing the moon at different phases (``new" and ``full").  The new moon, which in this context represents a crescent, is depicted in the lower--left of Frame 1, labelled as (h) while the full moon is the full circle depicted in the upper--left of Frame 2, labelled as (i).  Given that the cusps of the moon point toward the right, this stick seems to represent a waxing crescent moon, which is prominent in the early evening.  Lunar phases were a common method of determining time in message sticks (e.g. Howitt, 1898:317).

\section{Seasons \& Food Economics}

Marking the change of seasons was essential to the Aboriginal hunter/gatherer societies of Australia, and the positions of celestial bodies were a good way to accomplish this (see Table \ref{table:stars_seasons} for examples).  The change of seasons denoted altering weather patterns, the availability of particular food sources, and the breeding seasons of animals.  The various climactic and geographic regions of Australia have led Aboriginal groups to designate a wide variety of seasons.  In areas like northwest Victoria, Aboriginal groups had four distinct seasons (Stanbridge, 1858) while groups in Arnhem Land and many other areas had six (Thomson \& Peterson, 1983), with substantial variation across the continent.  In many communities, these seasonal changes are designated by the appearance of particular stars.

\begin{table}[h]
\caption{Examples of the significance to different Aboriginal communities of the heliacal (morning) rising of particular stars.  The star Parna is yet unidentified.}
\centering                          
\begin{tabular}{llllll}            
\hline
Group			&  	Location			&	Object					& 	Meaning					&	Reference\\
\hline
Boorong			& 	NE Victoria		&	Vega					&	Mallee Fowls Build Nests		&	Stanbridge (1858)\\
Kaurna			& 	Adelaide Region	&	Parna(?)					& 	Start of Autumn Rains		&	Gell (1842)\\
Pitjantjatjara		& 	Central Desert		&	Pleiades					& 	Dingo Breeding Season		&	Tindale (2005)\\
Warnindilyakwa	& 	Groote Eylandt		&	$\upsilon$,$\lambda$ Scor	& 	Start of Dry Season			&	Mountford (1956)\\
Yolngu			& 	Arnhem Land		&	Scorpius					& 	Arrival of Macassans		& 	Mountford (1956)\\
\hline                            
\end{tabular}
\label{table:stars_seasons}
\end{table}

\section{Observing Stellar Variability}
In his seminal work on Boorong ethnoastronomy, Stanbridge (1858) quoted a number of stars, clusters, planets, and celestial objects, citing their Boorong names.  For the entry {\it Collowgullouric War} (the wife of {\it War}, the Crow, pronounced ``Waah"), he wrote ``a large red star in Robur Carol, designated 966".  From this description, Hamacher \& Frew (2010) were able to show that this was a reference to the hyper--giant variable star Eta Carina during an eruptive period in the 1840s, when it became the second brightest star in the night sky after Sirius.  Not only had the Boorong noted this ``supernova impostor" event, but they incorporated it into their oral traditions as the wife of {\it War}.  This is the only definitive Indigenous record of Eta Carinae's eruption.

There is also evidence that more subtle variable stars were noted by Aboriginal Australians.  Daisy Bates from Ooldea, South Australia recorded that the local Aboriginal people had noted the variability of Betelgeuse in Orion (Fredrick, 2008).  According to an oral tradition, the stars that constitute Orion represent {\it Nyeeruna}, a hunter of women who chases the women of the Pleiades ({\it Yugarilya}).  Nyeeruna's right hand is represented by the star Betelgeuse ($\alpha$~Orionis), which holds a club he endeavours to fill with fire--magic to hurl at {\it Kambugudha}, the eldest sister of the {\it Yugarilya} (represented by the V--shape of the brighter stars of the Hyades in Taurus), who prevents Nyeeruna from ever reaching the Yugarilya, thus humiliating him.  In his rage, Nyeeruna reddens with fire and lust, but Kambugudha's magic and humiliation causes his fire magic to die out, becoming faint.  After some time, his magic comes back and his brightness increases.  Betelgeuse is a red super--giant star with noted semi--regular variability that ranges between 1.2 $<V_{mag}<$ --0.2 with a mean magnitude of 0.42.  These variations, first described by Herschel (1849), have a duration of several months to a year.

\section{Dynamics of the Earth--Moon--Sun System}
In most Aboriginal cultures, the moon is male and the sun is female (e.g. Haynes, 2000; Johnson, 1998; Fredrick, 2008; Norris \& Hamacher, 2009), although this is not universal (e.g. Meyer, 1846:11--12).  Some of the accounts describe the sun--woman as an aggressive lover that chases the moon--man, who avoids her advances by zig--zagging across his path in the sky.  This is an explanation of how the relative positions of the sun and moon change throughout the lunar month, where the moon can appear to the north, south, or in the same line as the sun.  The Yolngu have a tradition that on rare occasions, the sun--woman ({\it Walu}) manages to capture the moon--man ({\it Ngalindi}) and consummates their relationship before he manages to escape, explaining a solar eclipse (Warner, 1937:538).  A similar Dreaming from the Euahlayi of New South Wales says the sun--woman ({\it Yhi}) eclipsed the moon--man ({\it Bahloo}) in a jealous rage because he constantly avoided her advances.  Bahloo is saved by the intervention of celestial spirits (Parker, 1905:139-140; Reed, 1965:130).  A sexual encounter between the sun and moon as the cause of a solar eclipse is also found among the Wirangu of South Australia (Bates, 1944:211).  Other Aboriginal groups deduced that something covered the sun during an eclipse, although it was not always identified with the moon (see Hamacher \& Norris, 2011c).

Many oral traditions explain lunar phases in a cultural context.  For example, the full moon is a fat, lazy man called {\it Ngalindi} to the Yolngu of Arnhem Land.  His wives punish his laziness by chopping off bits of him with their axes, causing the waning moon.  He manages to escape by climbing a tall tree to follow the Sun, but is mortally wounded, and dies. After remaining dead for three days (new moon), he rises again, growing fat and round (waxing moon), until his wives attack him again in a cycle that repeats to this day (Wells, 1964; Hulley, 1996).  To coastal communities, the relationship between lunar phases and tides is well known.  According to the Yolngu and the Anindilyakwa (Mountford, 1956), when the tides are high, the water fills the moon as it rises at dusk (full moon).  As the tides drop, the moon empties (crescent) until the moon is high in the sky during dusk or dawn, at which time the tides fall and the moon runs out of water (first and third quarter).  The moon begins to fill again when it rises at dawn (new moon).

These accounts show that Aboriginal people paid careful attention to the motions of the sun and moon, explaining the mechanics of an eclipse and the relationship between lunar phases and ocean tides.

\section{Astronomical Measurements}
\begin{figure}[h]
\begin{center}
 \includegraphics[width=8cm]{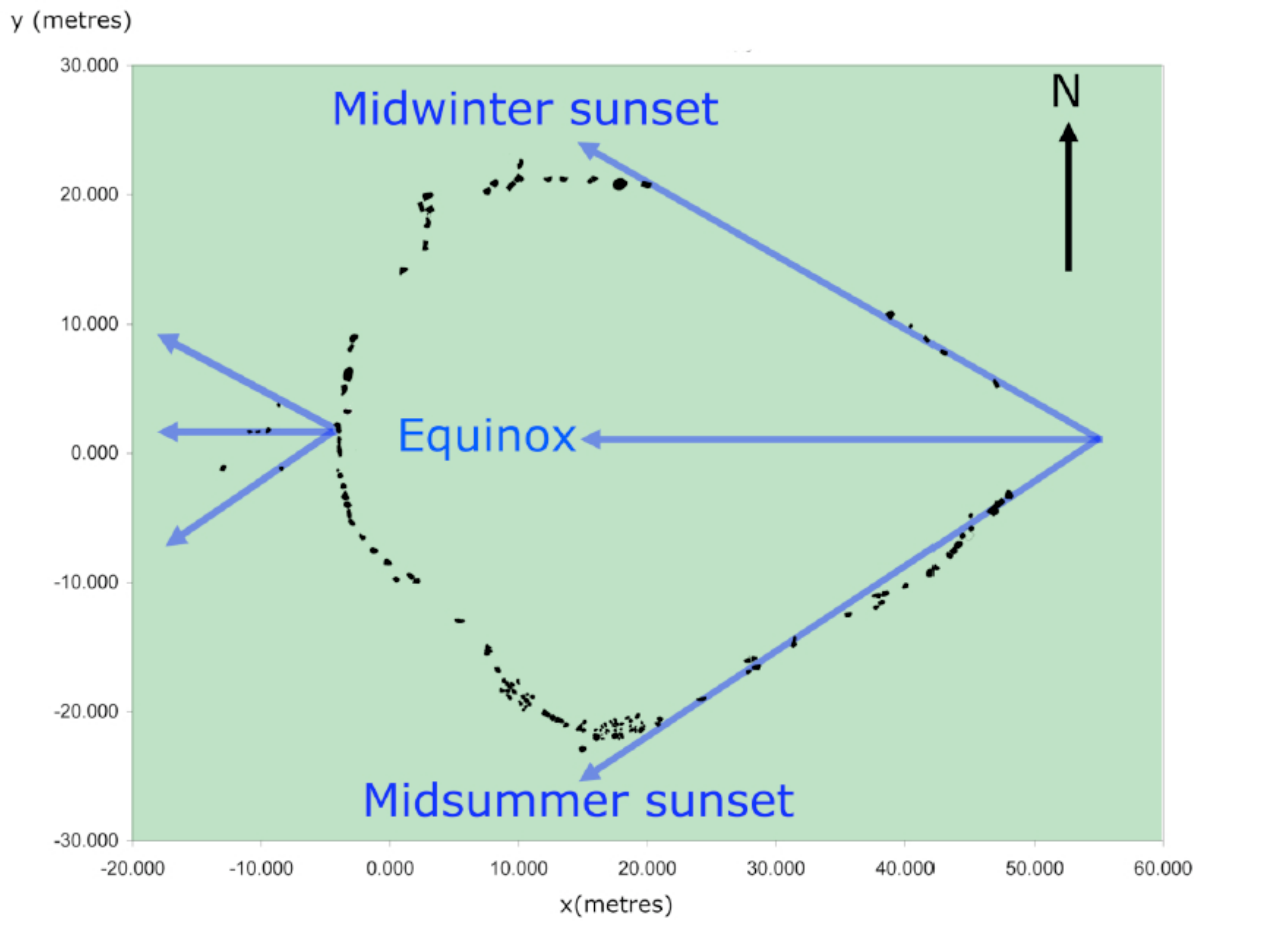} 
 \caption{The Wurdi Youang stone arrangement, which has two sets of alignments that indicate the setting sun at solstice and equinox.}
   \label{fig:wurdiyouang}
\end{center}
\end{figure}

The goal of this study is to determine whether Aboriginal people made astronomical observations and measurements.  Here, we discuss two examples from the archaeological record.  Wurdi Youang is an egg--shaped stone arrangement in Victoria, which was built by the Wathaurung people before European settlement. The stone arrangement is about 50 m in diameter with the major axis lying almost exactly East--West.  At its Western apex are three prominent waist--high stones. Morieson (2003) pointed out that outlying stones to the West of the circle, as viewed from these three stones, indicate the setting positions of the sun at the equinoxes and solstices. Norris et al (2011) have confirmed these alignments and have also shown that the straight sides of the circle also indicate the solstices and the three stones as viewed from the Eastern apex, define the setting sun at equinox (Fig. \ref{fig:wurdiyouang}).  This arrangement, if intended for this purpose, would have required careful observations of the sun throughout the year.  The age and exact purpose of this arrangement are unknown, but the two independent lines of evidence for solar indications support an astronomical relationship.

\begin{figure}[h]
\begin{center}
 \includegraphics[width=8cm]{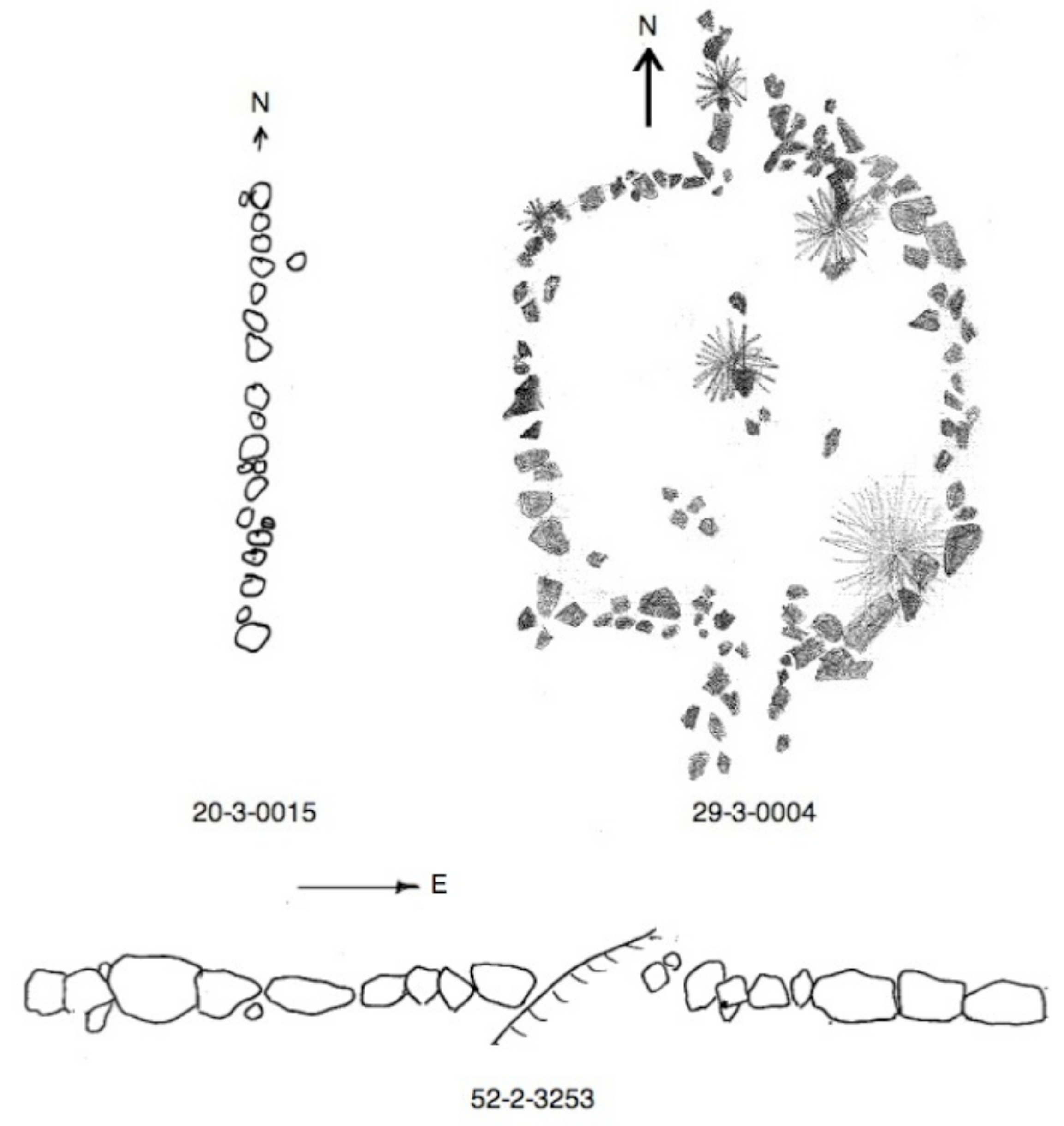} 
 \caption{Three examples of stone arrangements from the site cards that align to cardinal directions.  The number given is the Site Card ID.}
   \label{fig:stone_arrangements}
\end{center}
\end{figure}

\begin{figure}[h]
\begin{center}
 \includegraphics[width=12cm]{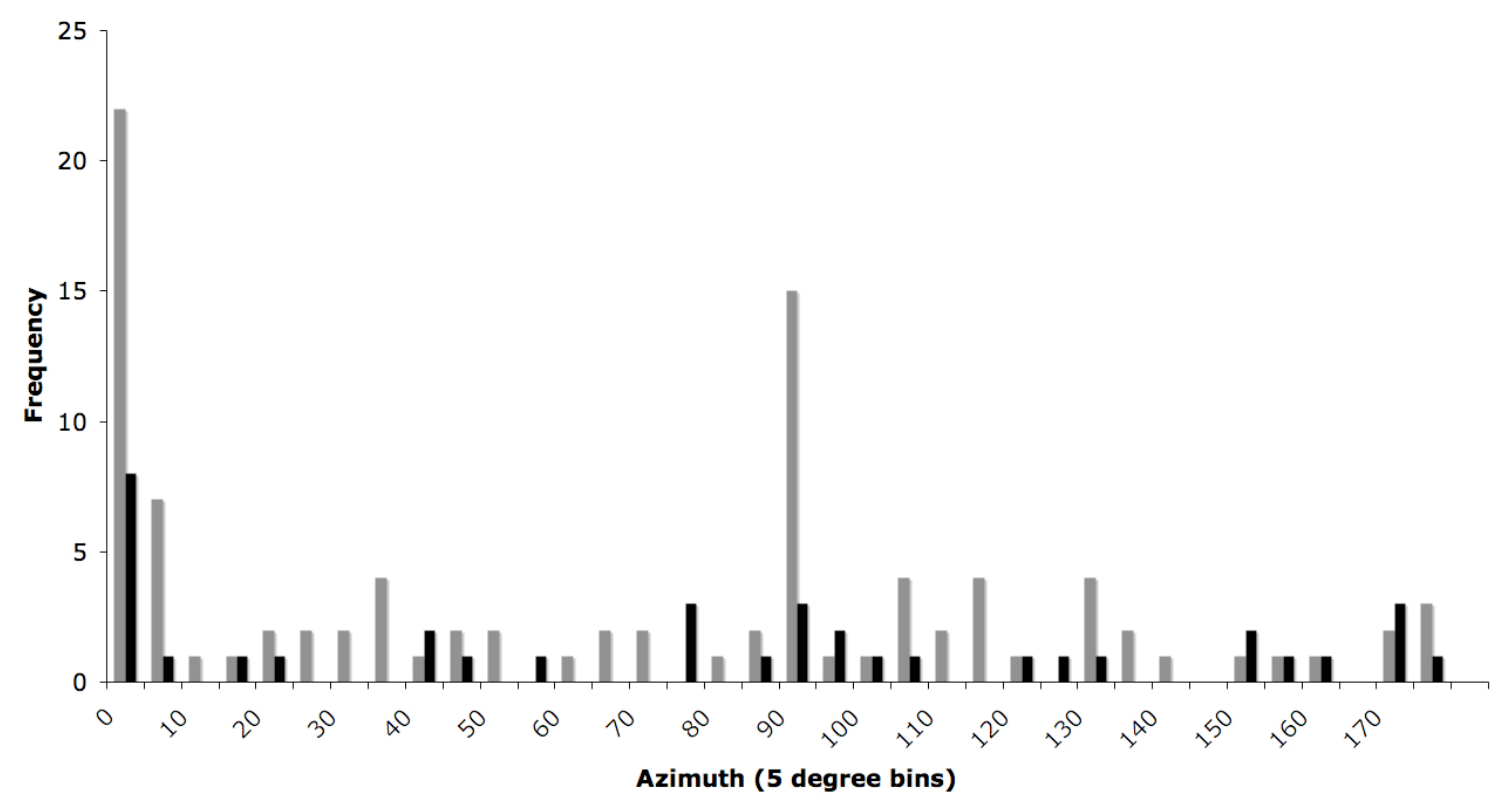} 
 \caption{Histogram of 134 azimuths in five--degree bins, categorised into 97 stone rows (grey) and 37 stone circles (black).  The histogram shows that stone rows are significantly more likely to be aligned to  cardinal directions, either North--South ($0^{\circ}$) or East--West ($90^{\circ}$) than other directions. Stone circles, on the other hand, show a preference for North--South but not East--West.}
   \label{fig:sta}
\end{center}
\end{figure}

Aboriginal stone arrangements in New South Wales have a variety of uses, including ceremonial and practical (e.g. McBryde, 1974; Bowdler, 1983; Attenbrow, 2002).  We present preliminary results of a cardinal alignment survey of stone arrangements in New South Wales.  We survey stone rows, pairs of stone circles/cairns (where the azimuth is measured from the centre of each circle or cairn), or single stone circles with an entrance and exit (Fig. \ref{fig:stone_arrangements}).  660 archaeological site cards were obtained from NSW Heritage, of which some contain a detailed archaeological survey.  We set selection criteria for determining the azimuth of these arrangements, using the site cards alone as an initial test.  We then measured the azimuths (labelling North as $0^{\circ}$, East as $90^{\circ}$, etc.) of each arrangement using these predetermined criteria, being careful not to bias the selection in favour of cardinal points.   Of the 660 site cards, we rejected 600 as either being ambiguous, having insufficient data, or being off--limits for cultural reasons.  From the 60 remaining sites, we obtained 134 orientations, although the site cards did not always specify whether the orientation was true or magnetic North.  However a preliminary analysis of the site card data shows (Fig. \ref{fig:sta}) a clear, statistically unambiguous preference toward cardinal directions. 

Specifically, a Monte Carlo simulation has shown that the probability of a peak as high as that at ($0^{\circ}$) in Fig. \ref{fig:sta} arising by chance alone is about 2 in 100 million, and the joint probability of getting both that and the peak at ($90^{\circ}$) is about $4 \times 10^{-12}$.  We conclude that this is not a chance occurrence, but that the stone rows are deliberately aligned on the cardinal points.  We are therefore conducting a series of field surveys of all sites with azimuths near cardinal directions to measure the accuracy with which these rows are aligned.

A preference towards cardinal azimuths indicates that the cardinal points were well known.  The determination of cardinal points requires careful measurements of the sun throughout the day and year, as the rising and setting position of the Sun and other bodies varies significantly through the year.

One technique to determine South is to observe the rotation of the Southern Cross over the course of a winter's night, marking the position on the horizon vertically below its extreme easterly and westerly positions, then marking the half--way point between them.  Similarly, East and West may be found by marking the extremes of the rising or setting sun's locations on the horizon over the year, then marking the halfway point between them.  Either technique requires a process of astronomical observation and measurement. Thus the very existence of a significant number of structures aligned to the cardinal points implies a degree of planning, observation, and measurement that seem to be absent from most anthropological accounts of Aboriginal cultures.  While any one structure might be aligned to the cardinal points purely by chance, we have shown that the probability of such a large number doing so is negligible.

\section{Conclusion}
We have given examples of how Aboriginal people made use of celestial phenomena for calendric, navigational, and cultural purposes and showed that Aboriginal people were careful observers of the night sky.   We have presented evidence that Aboriginal Australians noted the changing brightness of particular stars, deduced the complex motions of the sun and moon.  Even more significantly,  we have shown that Aboriginal Australians also oriented stone arrangements to cardinal directions and astronomical phenomena, such as the solstices and equinox.  This shows that Aboriginal Australians made careful observations and measurements of celestial bodies and positions, casting a new light on our understanding of the development of Aboriginal culture.

\begin{acknowledgments}
This work is dedicated to the hundreds of thousands of Indigenous Australians who lost their lives, land, and culture after the British colonisation of Australia.  Hamacher would like to thank the Department of Indigenous Studies at Macquarie University, the International Astronomical Union, the Astronomical Society of Australia, and the International Society of Archaeoastronomy \& Astronomy in Culture for funding to attend Oxford IX.  We also thank Dianne Johnson, John Morieson, John Clegg, Cilla Norris, Hugh Cairns, Bill Yidumduma Harney, Serena Fredrick, David Frew, Paul Curnow, Ros Haynes, NSW Heritage, and Kristina Everett.
\end{acknowledgments}

\end{document}